\documentclass[10pt,a4paper,twoside]{article}
\usepackage{epsfig}
\usepackage{baltlat6}
\usepackage{array}
\usepackage{here}
\def\be{\begin{equation}}
\def\ee{\end{equation}}
\def\ba{\begin{eqnarray}}
\def\ea{\end{eqnarray}}

\def\ltsima{$\; \buildrel < \over \sim \;$}
\def\simlt{\lower.5ex\hbox{\ltsima}}
\def\gtsima{$\; \buildrel > \over \sim \;$}
\def\simgt{\lower.5ex\hbox{\gtsima}}
\def\etal{{et al.\ }}

\usepackage[dvipsnames]{color}
\begin{document}
\ \
\vspace{0.5mm}
\setcounter{page}{277}

\titlehead{Baltic Astronomy, vol.???, , ????}

\titleb{Ionized gas velocity dispersion and multiple supernova explosions}

\begin{authorl}
\authorb{Evgenii O. Vasiliev}{1,2},
\authorb{Alexei V. Moiseev}{3,4} and 
% \authorb{Biman B. Nath}{5} and \break
\authorb{Yuri A. Shchekinov}{2}
\end{authorl}

\begin{addressl}
\addressb{1}{Institute of Physics, Southern Federal University, Stachki Ave. 194, Rostov-on-Don, 344090 Russia;
eugstar@mail.ru}
\addressb{2}{Department of Physics, Southern Federal University, Sorge Str. 5, Rostov-on-Don, 344090 Russia}
\addressb{3}{Special Astrophysical Observatory, RAS, Nizhnii Arkhyz, Karachaevo-Cherkesskaya Republic,  369167 Russia}
\addressb{4}{Sternberg Astronomical Institute, Moscow M.V. Lomonosov State University, Universitetskij pr., 13, 119992 Moscow, Russia}
% \addressb{5}{Raman Research Institute, Sadashiva Nagar, Bangalore 560080, India}
\end{addressl}

\submitb{Received: 20?? December 2; accepted: 20?? December 15}

\begin{summary} 
Using 3D numerical simulations we study the evolution of the H$\alpha$ intensity and velocity dispersion for single 
and multiple supenova (SN) explosions. We find that the $I_{\rm H\alpha}-\sigma$ diagram obtained for simulated gas 
flows is similar in shape to that observed in dwarf galaxies. We conclude that colliding SN shells with significant
difference in age are resposible for high velocity dispersion that reaches values high as $\simgt 100$~km~s$^{-1}$. 
Such a high velocity dispersion could be hardly got for a single SN remnant. Peaks of velocity dispersion on the 
$I_{\rm H\alpha}-\sigma$ diagram may correspond to several stand-alone or merged SN remnants with moderately 
different ages. The procedure of the spatial resolution degrading in the H$\alpha$ intensity and velocity dispersion 
maps makes the simulated  $I_{\rm H\alpha}-\sigma$ diagrams close to those observed in dwarf galaxies not only in 
shape, but also quantitatively.
\end{summary}

\begin{keywords} 
galaxies: ISM -- ISM: bubbles -- shock waves -- supernova remnants -- kinematics and dynamics 
\end{keywords}

%% \resthead is the RUNNING TITLE at top of the pages
\resthead{$I_{\rm H\alpha}-\sigma$ relation and multiple SNe}
{E.O. Vasiliev \etal}

\sectionb{1}{INTRODUCTION}

%%%%%%%%%%%%%%%%%%%%%%%%%%%%%%%%%%%%%%%%%%%%%%%%%%%%%%%%%%%%%%%%%%
\begin{figure}[!tH]
\vbox{
\centerline{
\psfig{figure=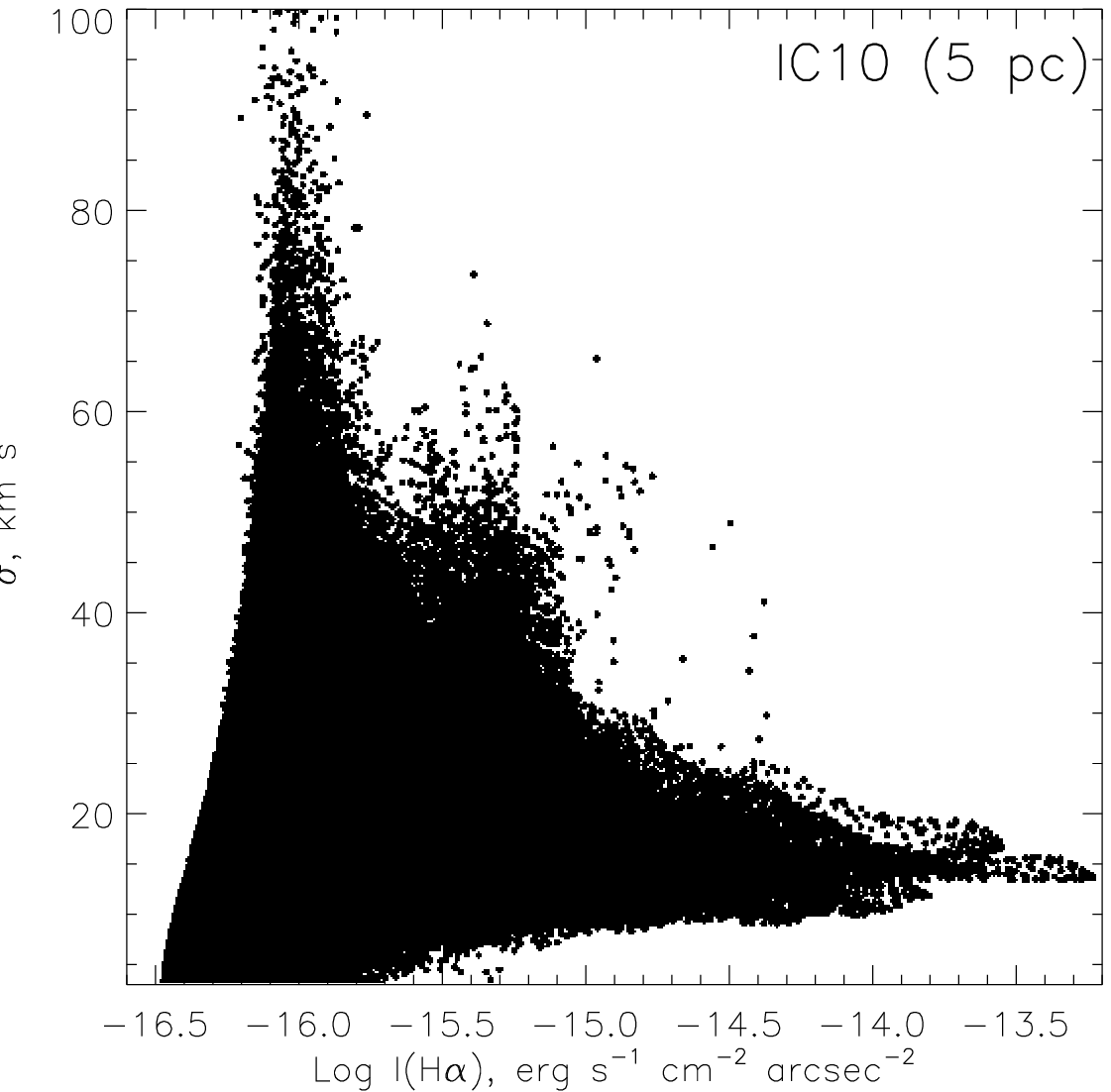,width=50mm,angle=0,clip=}
\psfig{figure=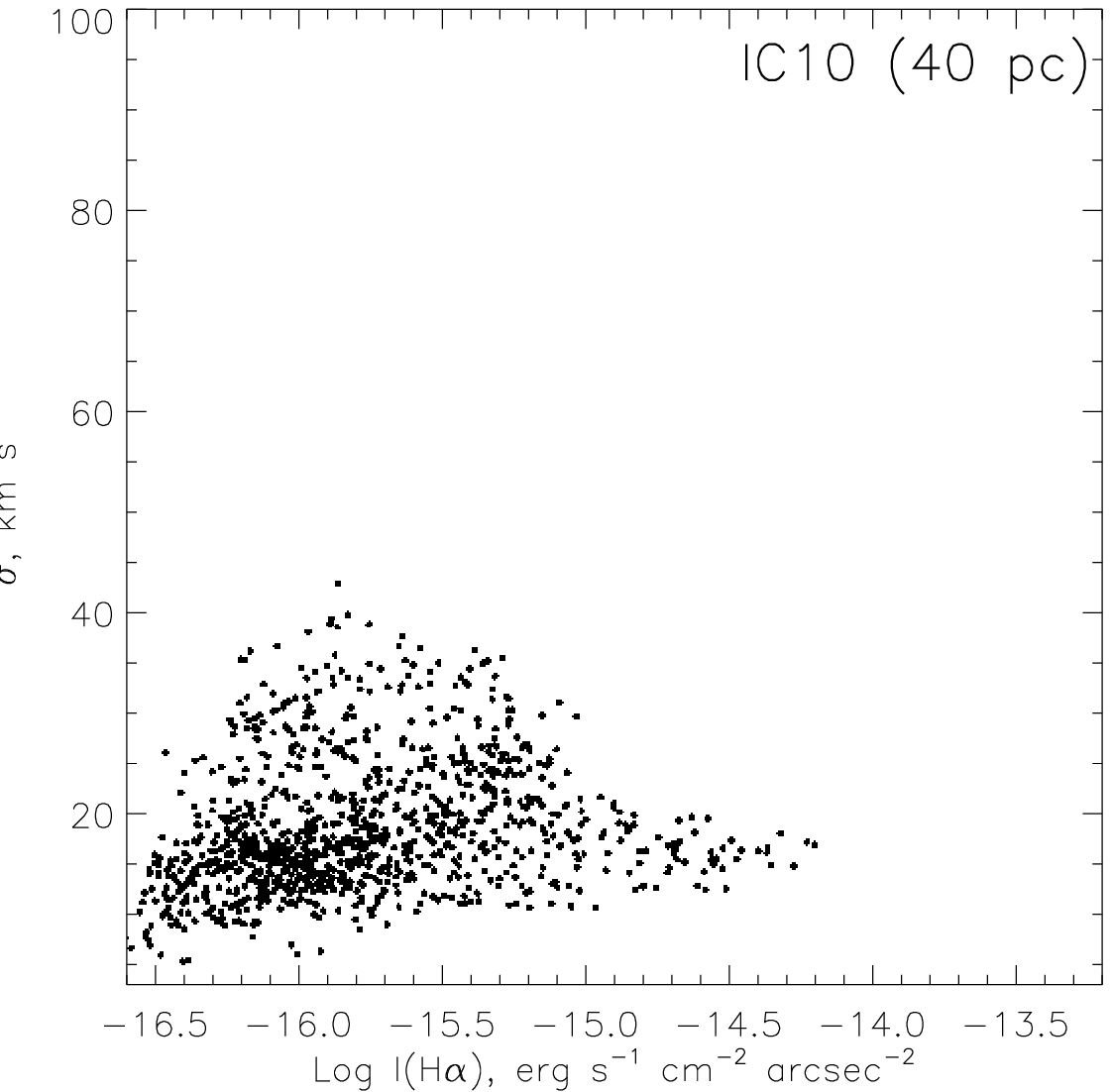,width=50mm,angle=0,clip=}
}
\break
\centerline{
\psfig{figure=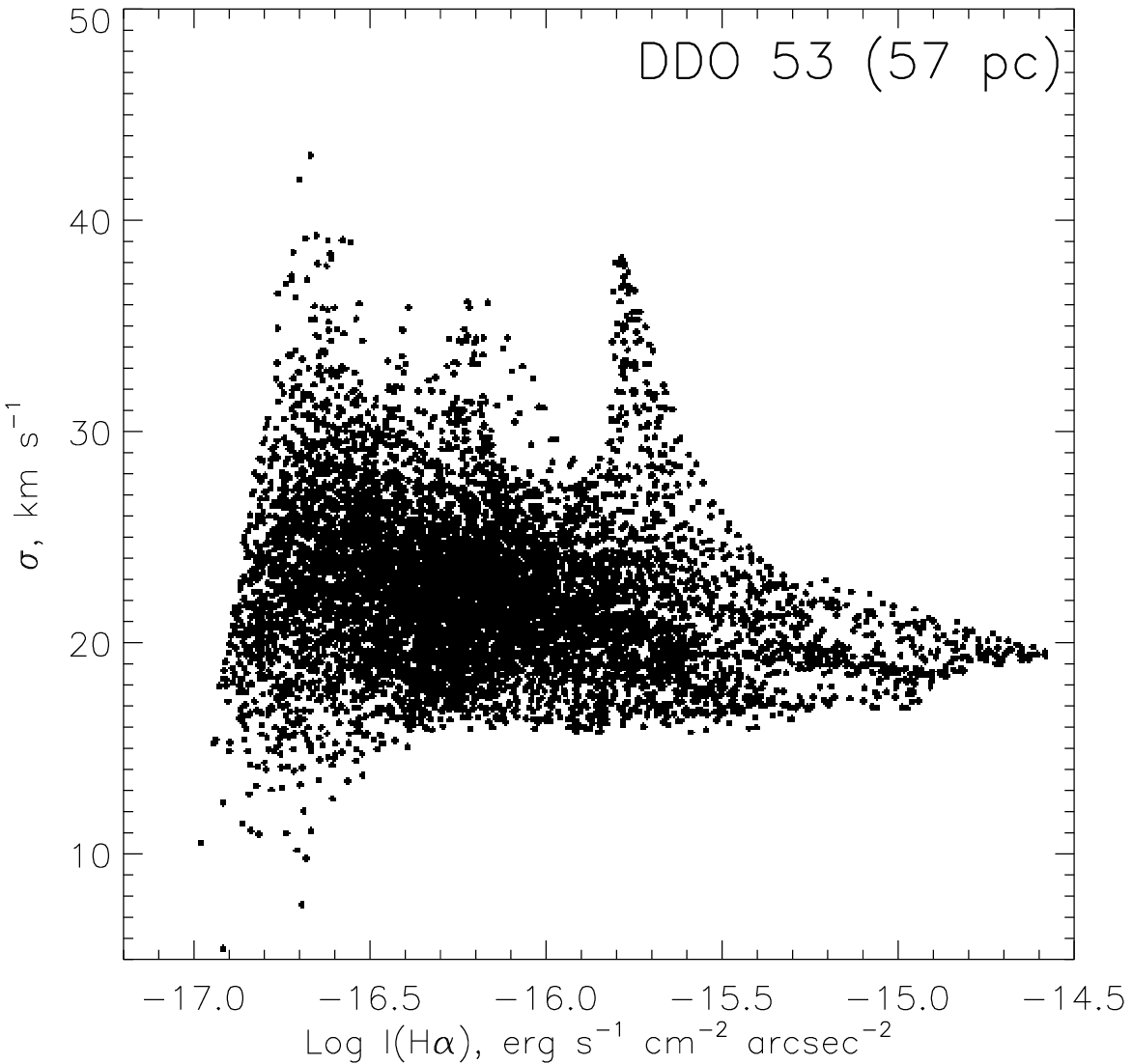,width=50mm,angle=0,clip=}
\psfig{figure=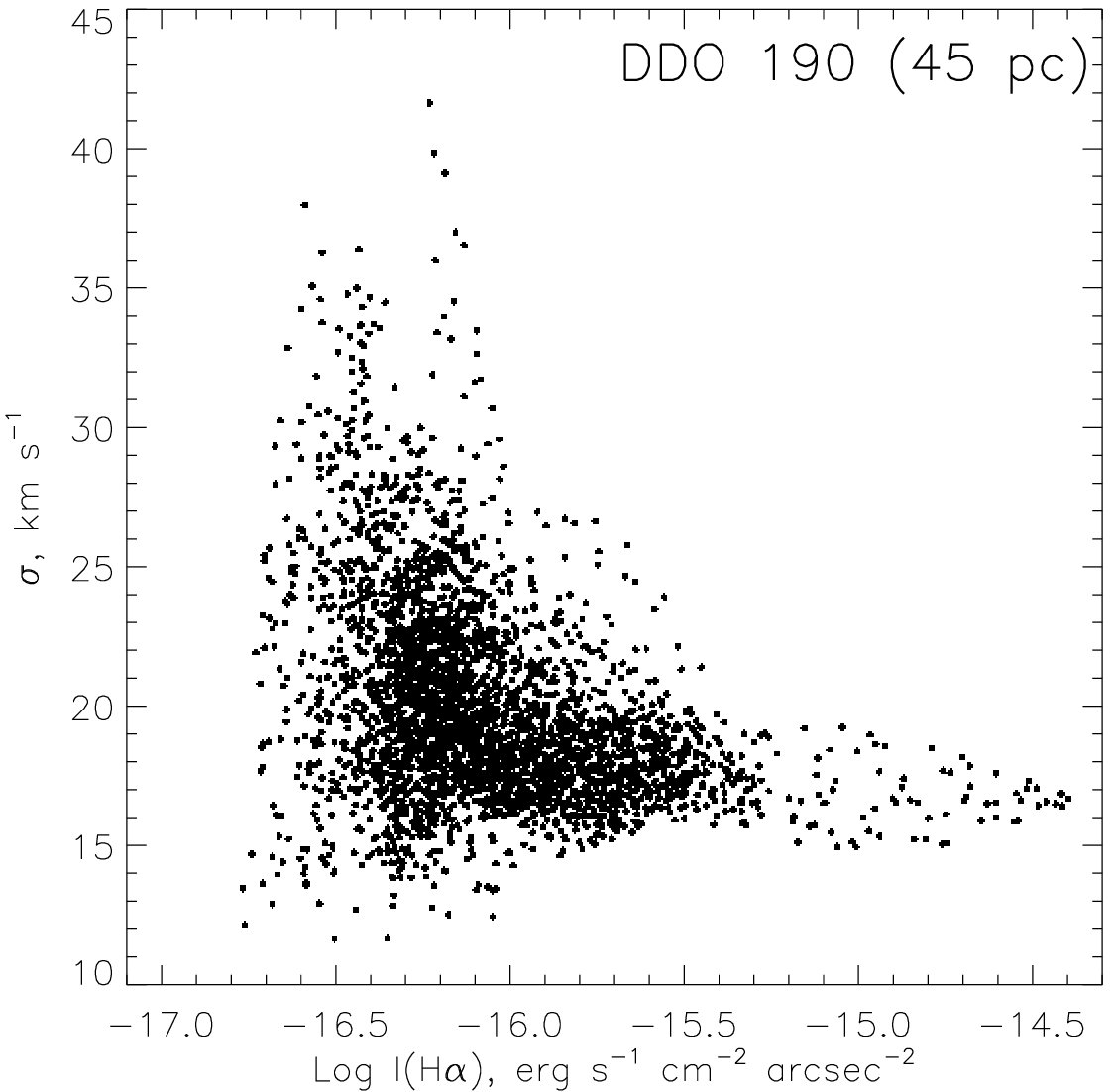,width=50mm,angle=0,clip=}
}
\vspace{1mm}
\captionb{1}
{
The $I_{\rm H\alpha}-\sigma$ diagram for nearby dwarf galaxies according Moiseev \& Lozinskaya (2012): IC~10 (original and smoothed
data), DDO~53 and DDO~190. The spatial resolution is marked in the brackets.
}
}
\end{figure}
%%%%%%%%%%%%%%%%%%%%%%%%%%%%%%%%%%%%%%%%%%%%%%%%%%%%%%%%%%%%%%%%%%

The 3D spectroscopy in optical emission lines yields a detailed information about the kinematics of the ionized gas in the 
interstellar medium (ISM) of external galaxies with a seeing-limited angular resolution. On one hand, it provides the angular 
resolution $1-3$~arcsec, that is an order of magnitude higher than that reached in radio observations. But on the other hand, 
it is limited by only to the regions around ionizing radiation sources, like OB stars, stellar winds and supernova explosions. 
Kinetic energy output from these  objects may be responsible for driving turbulent motions in ambient gas and forming galactic 
outflows. The observed characteristic of these motions is a velocity dispersion $\sigma$ determinated as a standard deviation 
of the Gaussian profile describing  Balmer emission lines after accounting for the instrumental effects and subtracting the 
contribution of the natural and thermal broadening in the HII regions. For understanding observations of starforming complexes
Mu\~noz-Tu\~n\'on \etal (1996) and Yang \etal (1996) have proposed to use the relation between the H$\alpha$ intensity (i.e. the 
emission line surface brightness) and velocity dispersion -- the $I_{\rm H\alpha}-\sigma$ diagram. More recently Mart\'inez-Delgado
\etal (2007), Bordalo \etal (2009), Moiseev \etal (2010), Moiseev \& Lozinskaya (2012) have considered $I_{\rm H\alpha}-\sigma$ 
diagrams to study the ionized gas in several dwarf galaxies. Figure~1 presents such diagrams for nearby dwarf irregular galaxies 
IC~10, DDO~190 and DDO~53. The observational data for these diagrams were obtained using a scanning Fabry-Perot interferometer 
with the 6-m telescope of the SAO RAS and described in details by Moiseev \& Lozinskaya (2012). The spatial resolution in the
observations of the nearest galaxy IC~10 (5~pc) is 8 times better than that for DDO~190 and DDO~53 galaxies. In observations with 
lower resolution gaseous motions in small HII regions, thin shells and colliding shocks are averaged with neighbouring low-velocity
flows. So the data on the small-scale kinematics of gas is lost for two latter galaxies that are more distant than IC~10. To imitate 
the effect of low resolution, the original data cube for IC 10 was first smoothed by the two-dimensional Gaussians, and then binned, 
so that the resulting pixel size compared to that for DDO~190 and DDO~53 (Moiseev \& Lozinskaya 2012). The diagram for IC~10 after
degrading resolution looks similar to these for DDO~190 and DDO~53 and fills the same range of $\sigma$. In general, all such 
diagrams have similar shape, because there are several main sources that contribute to H$\alpha$ line intensity and that drive 
motions in turbulent ISM. Namely, there are HII regions, stellar winds and supenova remnants. Acting together they form multiphase
turbulent ISM. Up-to-date only quantitive interpretation of $I_{\rm H\alpha}-\sigma$ diagrams were proposed by Mu\~noz-Tu\~n\'on 
\etal (1996). However one can be interested in how ionized gas has reached to the state, which corresponds to shape of $I_{\rm
H\alpha}-\sigma$ diagrams observed in galaxies. 

In this note we consider the evolution of the H$\alpha$ intensity and velocity dispersion for single and multiple supenova explosions.

\sectionb{2}{NUMERICAL MODEL}

We are interested in the relation between emissivity in H$\alpha$ line and velocity dispersion for gaseous flows driven 
by multiple SN explosions in starforming galaxies (the detailed study of multiple SNe dynamics can be  found in Vasiliev
\etal 2014). We follow the evolution of the H$\alpha$ emission intensity and velocity dispersion in a small box of
200~pc$^3$. For SN rate we consider a finite number of SNe exploded each $10^5$ and $10^4$~yrs, which corresponds to 
the rate $1.25\times 10^{-12}$ and $1.25\times 10^{-11}$~pc$^{-3}$~Myr$^{-1}$ in case of continous starformation on 
galactic scales. For comparison we investigate how similar values evolve in case of single SN explosion.

We carry out 3-D hydrodynamic simulations (Cartesian geometry) of multiple/single SNe explosions. We take periodic 
boundary conditions. The computational domain has size $200^3$ pc$^3$, which has $300^3$ cells, corresponding to a 
physical cell size of $0.75$ pc. The background number density considered are $0.1$ and $1$~cm$^{-3}$, and the background
temperature is $10^4$~K. The metallicity is constant within the computational domain (we do not consider here the mixing 
of metals ejected by SNe, this question will be studied in a separate work), and we consider cases with $Z=0.1, 1$ Z$_\odot$. 
We inject the energy of each SN in the form of thermal energy in a region of radius $r_i=1.5$ pc. The energy of each
SN equals $10^{51}$~erg. SNe are distributed uniformly and randomly over the computational domain. 

We use three-dimensional unsplit total variation diminishing (TVD) code based on the Monotonic Upstream-Centered 
Scheme for Conservation Laws (MUSCL)-Hancock scheme and the Haarten-Lax-van Leer-Contact (HLLC) method (e.g. Toro 
1999) as approximate Riemann solver. This code has successfully passed the whole set of tests proposed in (Klingenberg 
\etal 2007).

In the energy equation we take into account cooling processes adopted the tabulated non-equilibrium cooling curve 
(Vasiliev 2013). This cooling rate is obtained for a gas cooled isobarically from $10^8$ down to 10~K. 
The full description of our method of cooling rate calculations and the references to the atomic data can be found 
in (Vasiliev 2011, 2013). Briefly, the non-equilibrium calculation includes the ionization kinetics of all ionization 
states for the following chemical elements: H, He, C, N, O, Ne, Mg, Si, Fe as well as molecular hydrogen kinetics at 
$T<10^4$~K. The heating rate is adopted to be constant, whose value is chosen so that the background gas does not cool. 
The stabilization vanishes when the density and temperature goes out of the narrow range near the equilibrium state.

We use the pre-computed cooling rates because the self-consistent calculation of cooling rates in multi-dimensional 
dynamics is a very time consuming task. In general, the evolution of gas behind shock waves with velocities higher 
than $\simgt 150$~km~s$^{-1}$ are very close to that of a gas cooled from very high temperature $T=10^8$~K (e.g.,
Vasiliev 2012). Here we study multiple SN explosions, whose shells collide and merge with each other with typical 
velocities higher than 100~km~s$^{-1}$.Therefore the non-equilibrium cooling rates can be pre-computed for a gas 
cooled from very high temperature, e.g. $T=10^8$~K and these rates can be used to study SN shell evolution in 
tabulated form.

\sectionb{3}{RESULTS}

%%%%%%%%%%%%%%%%%%%%%%%%%%%%%%%%%%%%%%%%%%%%%%%%%%%%%%%%%%%%%%%%%%
\begin{figure}[!tH]
\vbox{
\centerline{
\psfig{figure=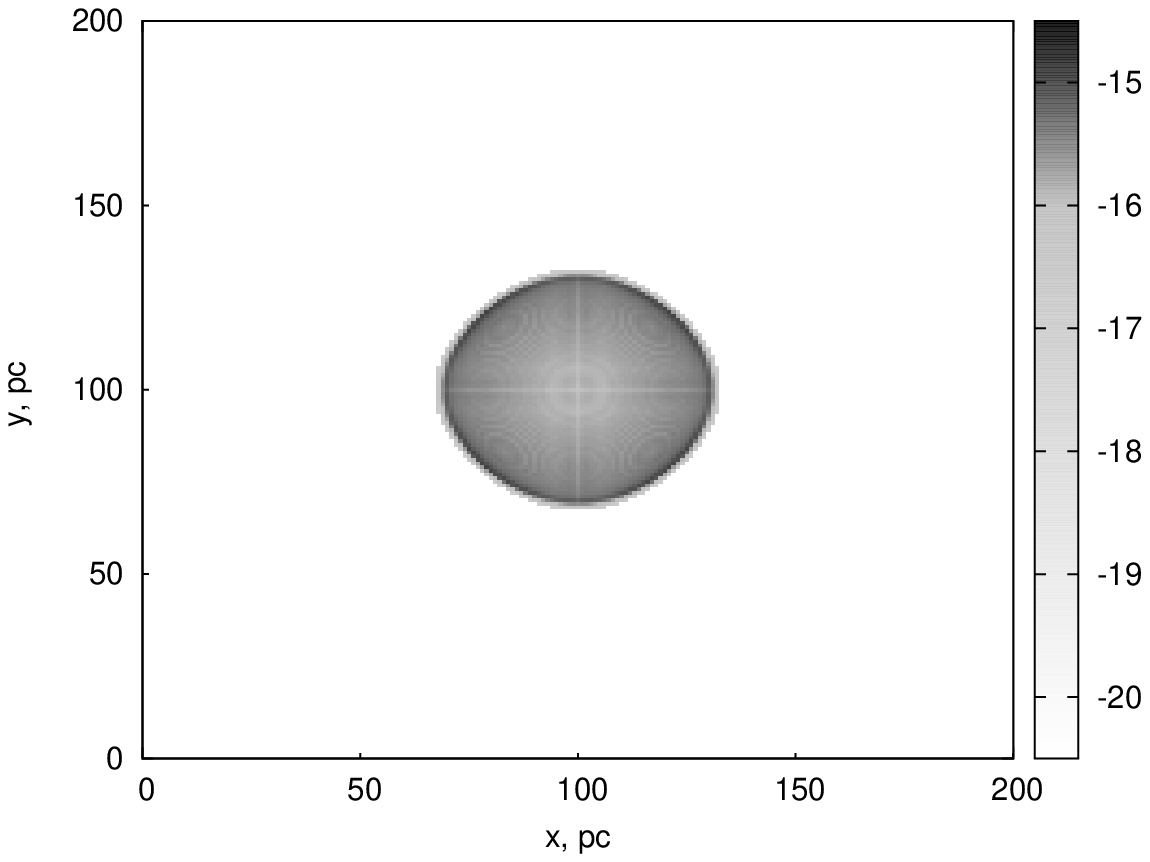,width=40mm,angle=0,clip=}
\psfig{figure=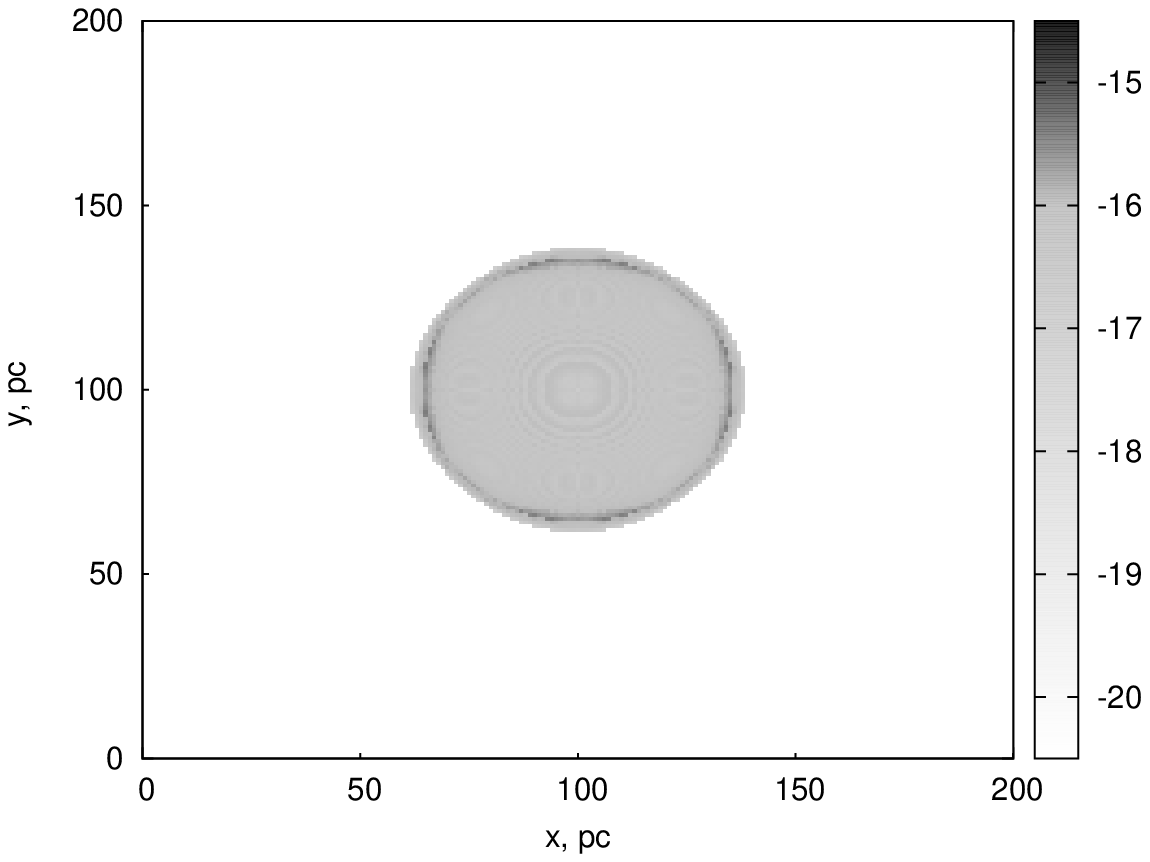,width=40mm,angle=0,clip=}
\psfig{figure=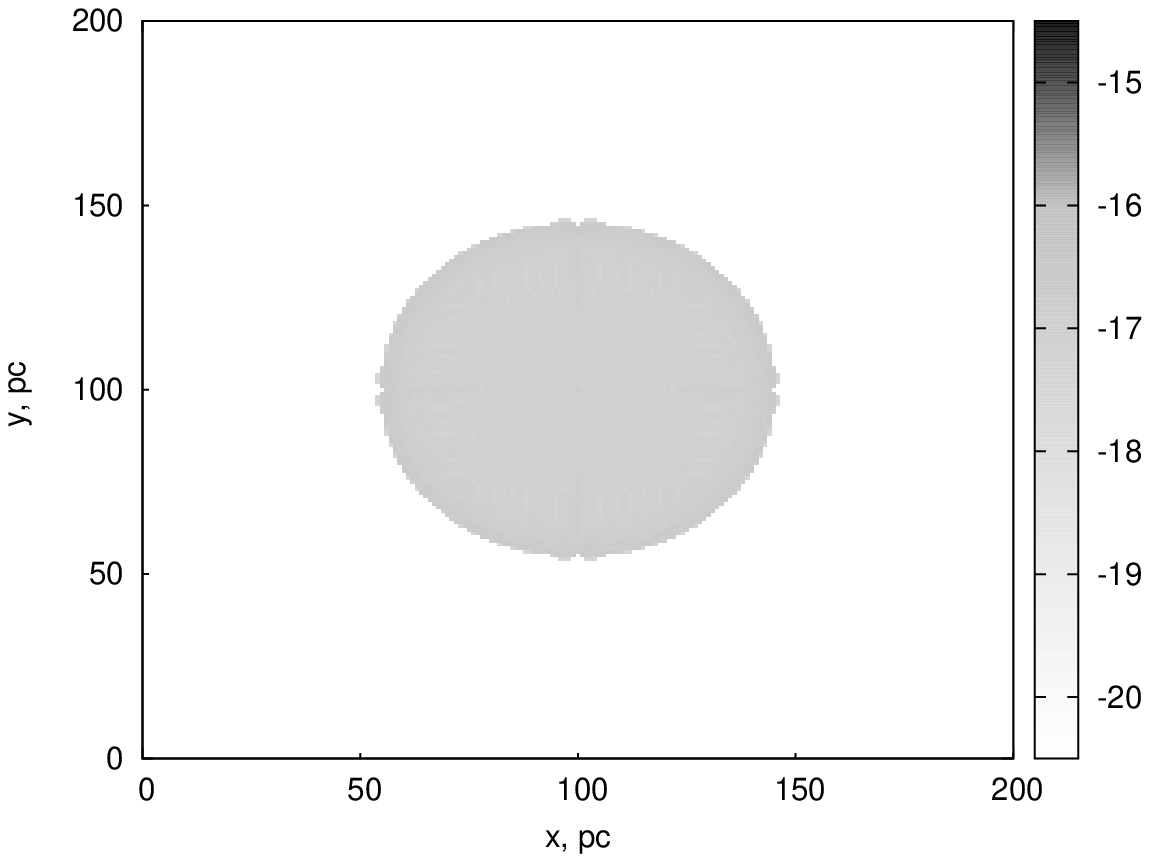,width=40mm,angle=0,clip=}
}
\break
\centerline{
\psfig{figure=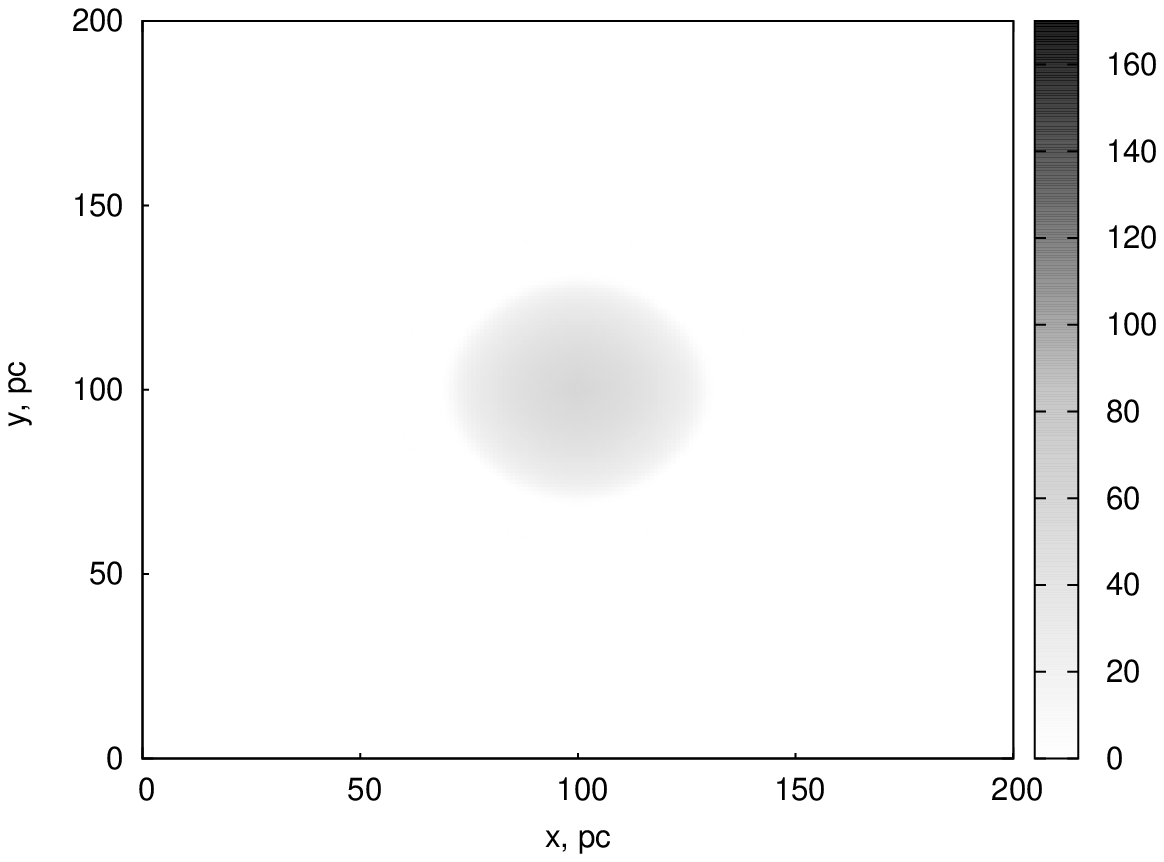,width=40mm,angle=0,clip=}
\psfig{figure=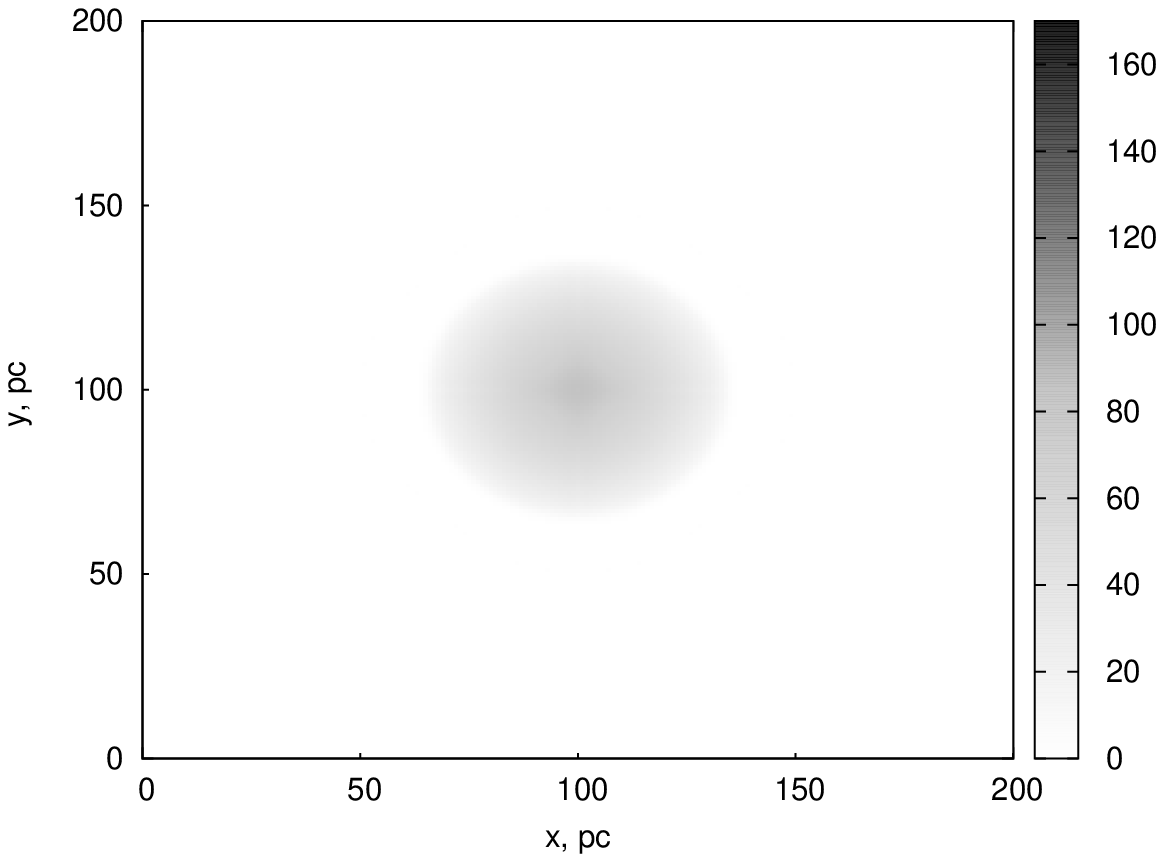,width=40mm,angle=0,clip=}
\psfig{figure=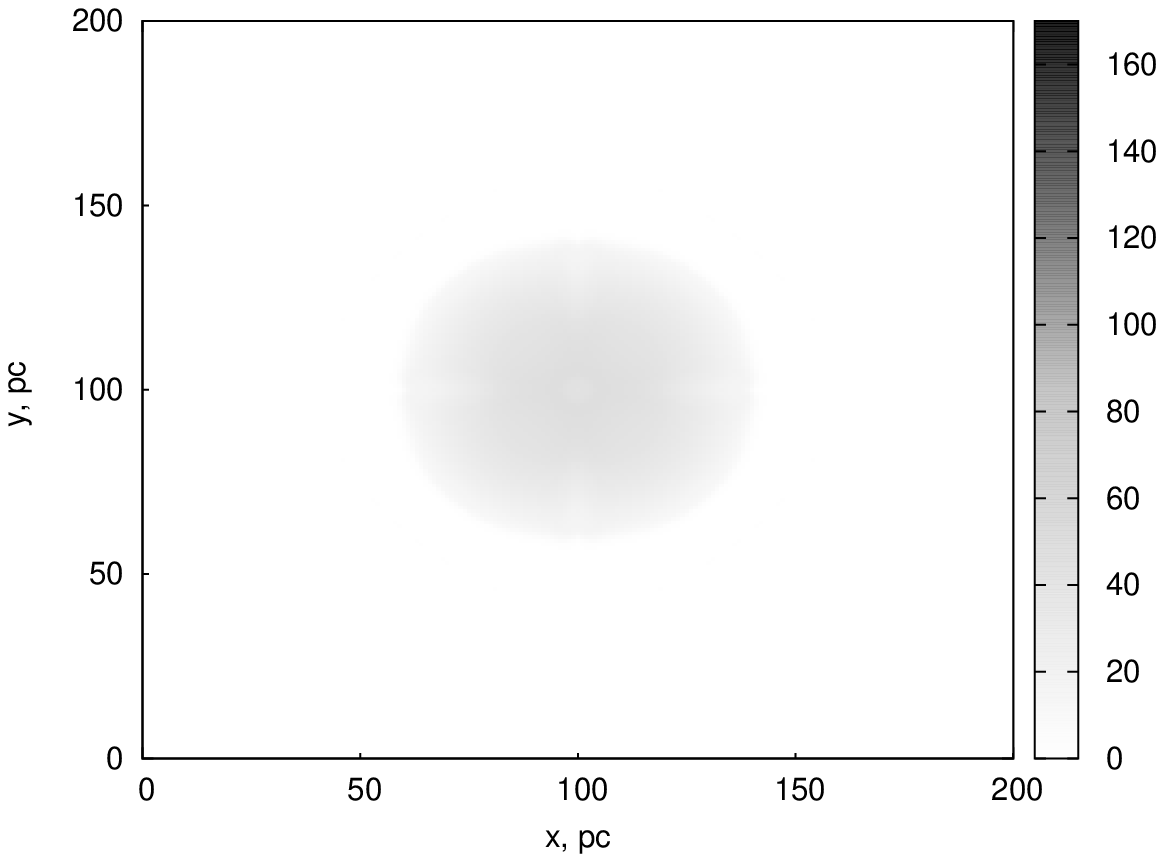,width=40mm,angle=0,clip=}
}
\vspace{1mm}
\captionb{2}
{
The H$\alpha$ intensity $I_{\rm H\alpha}$ (in erg~cm$^{-2}$~s$^{-1}$~arcsec$^{-2}$) -- top row, velocity dispersion 
(in km/s) maps --  bottom row, for single SN explosion at $t=0.1$, 0.2 and 0.4~Myr (from left to right columns). 
}
}
\label{ha-maps-1sn}
\end{figure}
%%%%%%%%%%%%%%%%%%%%%%%%%%%%%%%%%%%%%%%%%%%%%%%%%%%%%%%%%%%%%%%%%%

%%%%%%%%%%%%%%%%%%%%%%%%%%%%%%%%%%%%%%%%%%%%%%%%%%%%%%%%%%%%%%%%%%
\begin{figure}[!tH]
\vbox{
\centerline{
\psfig{figure=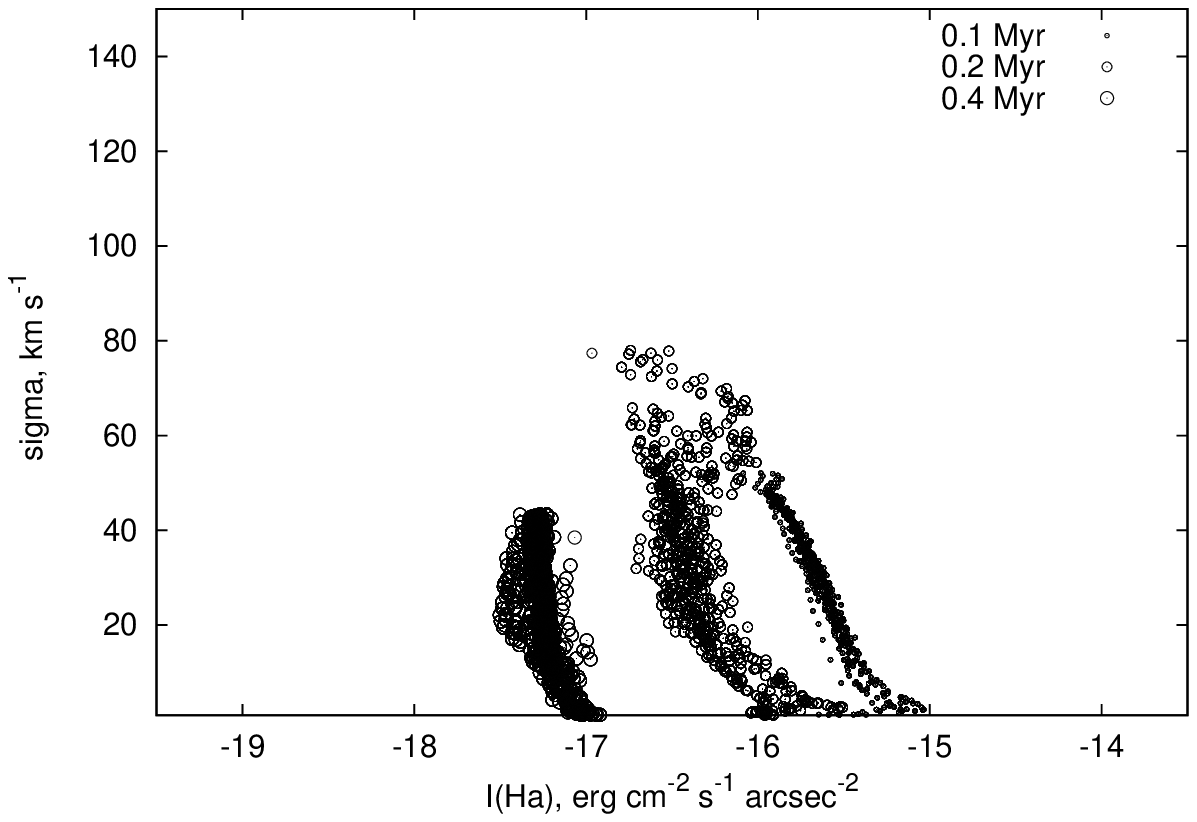,width=80mm,angle=0,clip=}
}
\vspace{1mm}
\captionb{3}
{
The $I_{\rm H\alpha}-\sigma$ diagram for single SN explosion at $t=0.1$, 0.2 and 0.4~Myr (from small to large symbols). 
}
}
\label{ha-dia-1sn}
\end{figure}
%%%%%%%%%%%%%%%%%%%%%%%%%%%%%%%%%%%%%%%%%%%%%%%%%%%%%%%%%%%%%%%%%%

First of all to understand H$\alpha$ emission and velocity dispersion distributions for multi-SNe explosions we consider
these distributions for single SN exploded in homogeneous medium. Figure~2 shows the H$\alpha$ intensity and velocity
dispersion maps for three time moments. One can see the drop of the average intensity with time, and note that the 
velocity dispersion in the direction of the explosion center grows at $t\sim 0.2$~Myr and than decreases. Around $t\sim
0.2$~Myr the reverse shock overtakes the blastwave due to decelerating the latter. Because of higher velocity of the 
reverse shock the velocity dispersion increases. After short interaction between shocks  the dispersion drops monotonically 
in time. Figure~3 shows the $I_{\rm H\alpha}-\sigma$ diagram for single SN explosion. It is clearly seen that
the locus of points corresponded to the same time moment shifts with time to region with lower H$\alpha$ intensity. The
maximum velocity dispersion is about 80~km~s$^{-1}$ reaches when the reverse shock is close to the blastwave. For older
remnant the intensity and velocity dispersion  becomes smaller. Decrease of the ambient density leads to longer evolution, 
but the maximum velocity dispersion after the beginnig of the radiation phase depends on the ambient density value slightly.

%%%%%%%%%%%%%%%%%%%%%%%%%%%%%%%%%%%%%%%%%%%%%%%%%%%%%%%%%%%%%%%%%%
\begin{figure}[!tH]
\vbox{
\centerline{
\psfig{figure=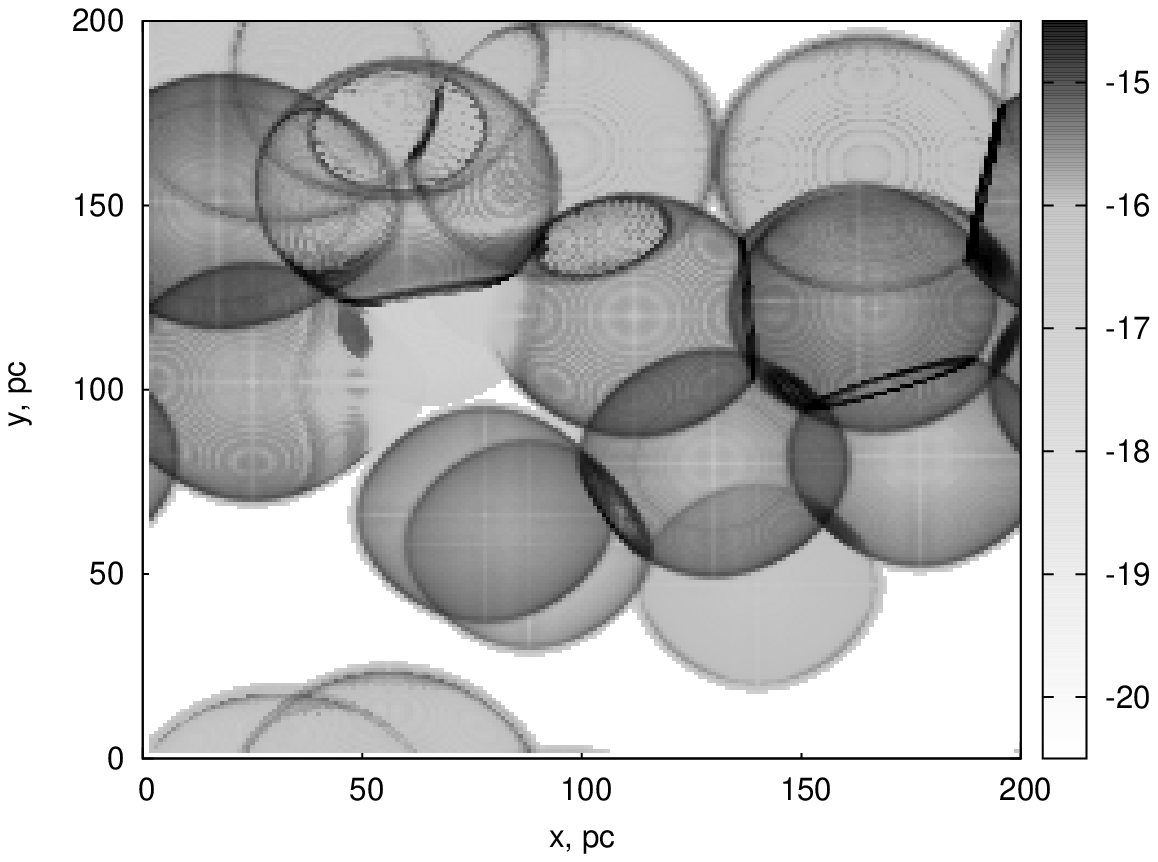,width=40mm,angle=0,clip=}
\psfig{figure=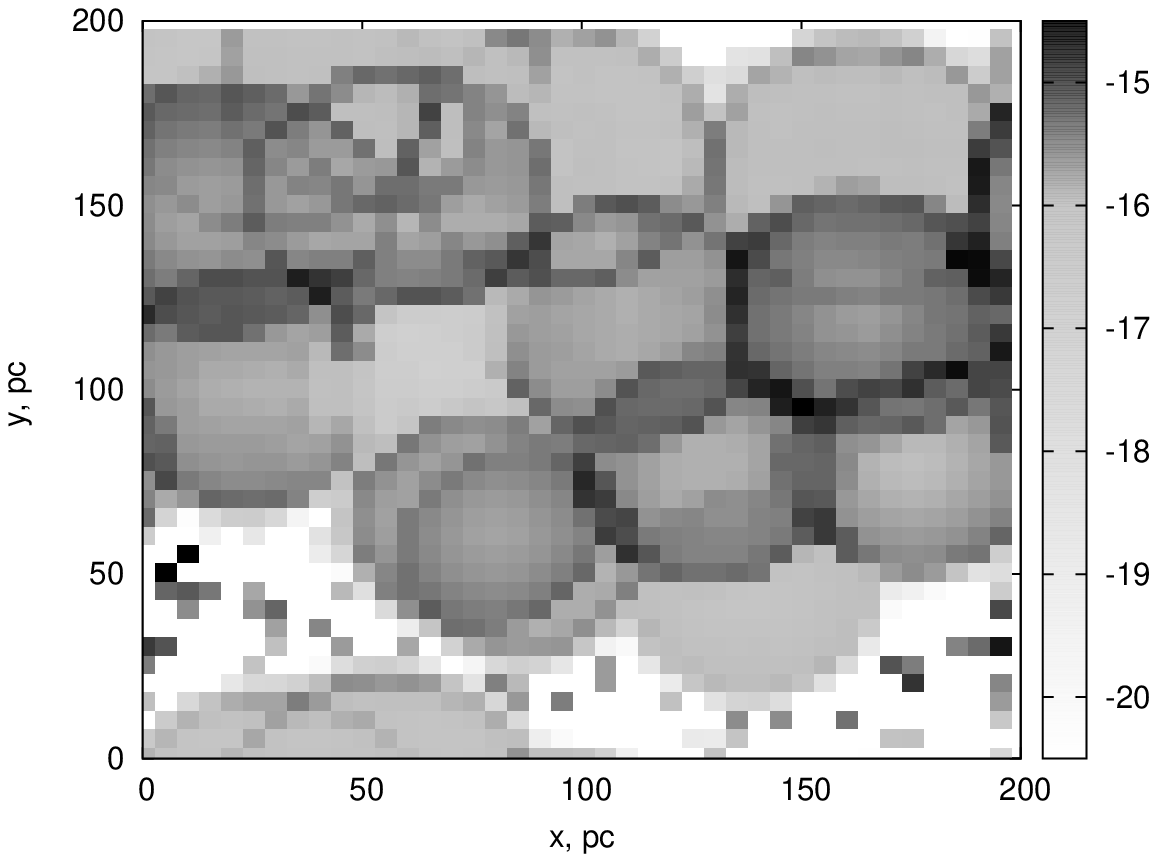,width=40mm,angle=0,clip=}
\psfig{figure=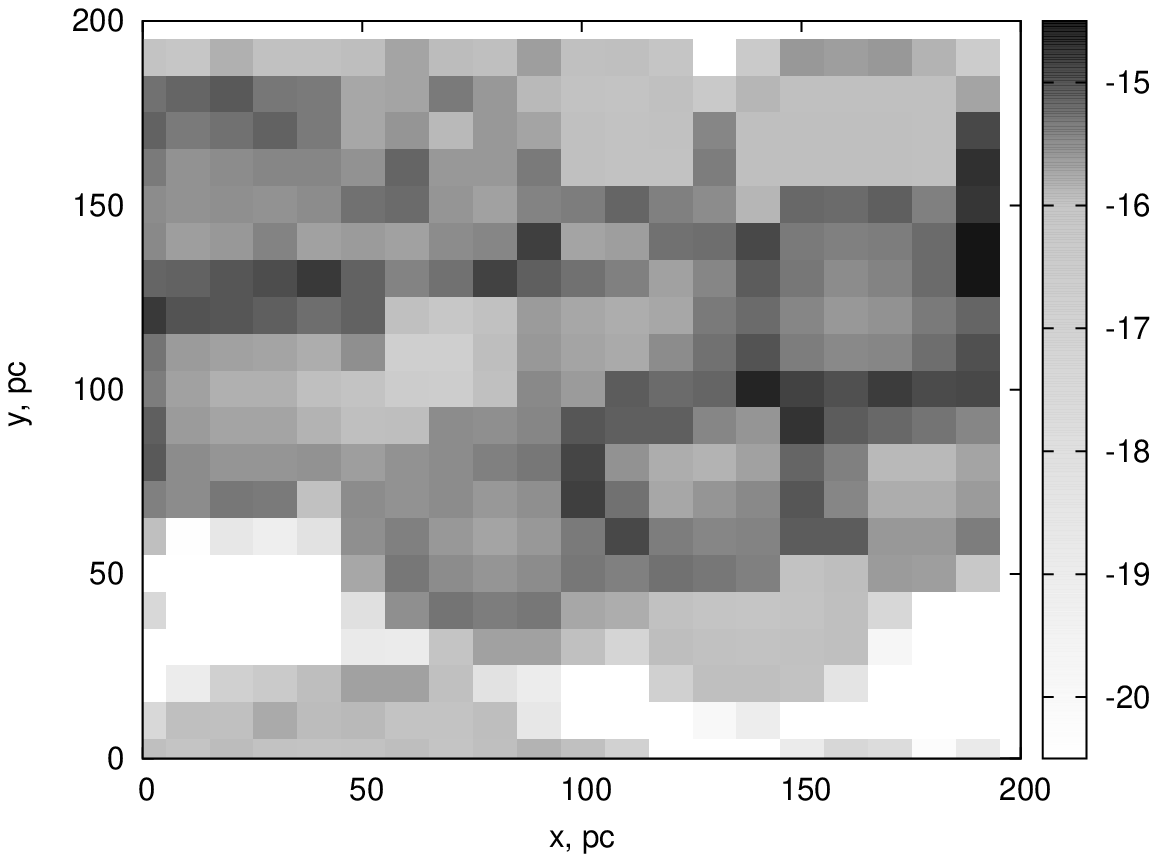,width=40mm,angle=0,clip=}
}
\break
\centerline{
\psfig{figure=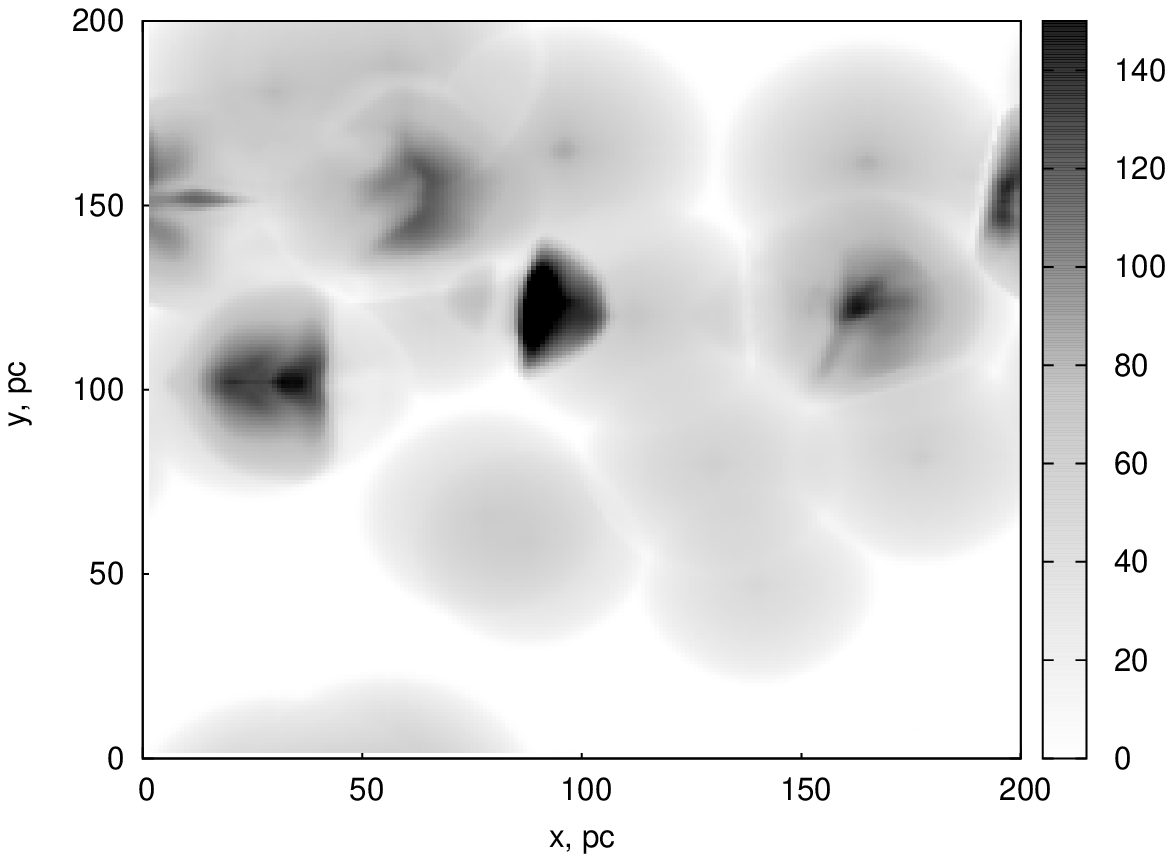,width=40mm,angle=0,clip=}
\psfig{figure=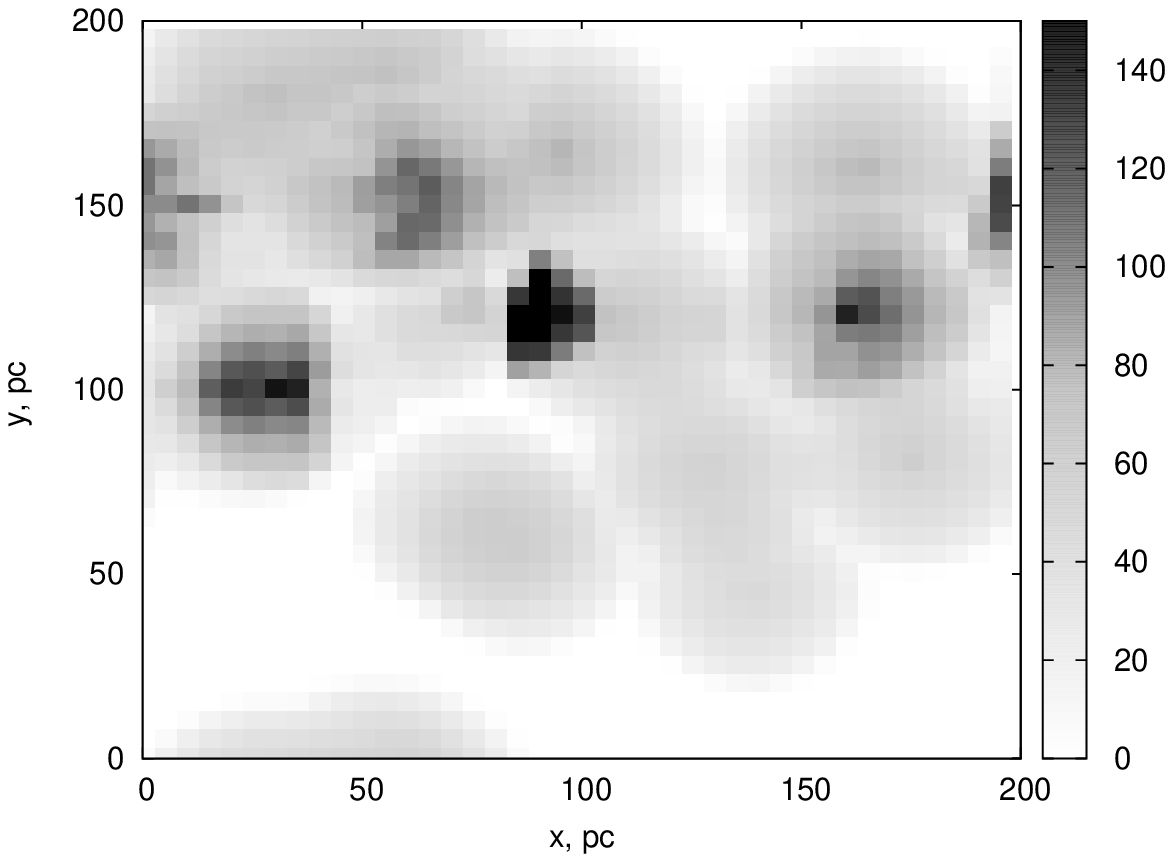,width=40mm,angle=0,clip=}
\psfig{figure=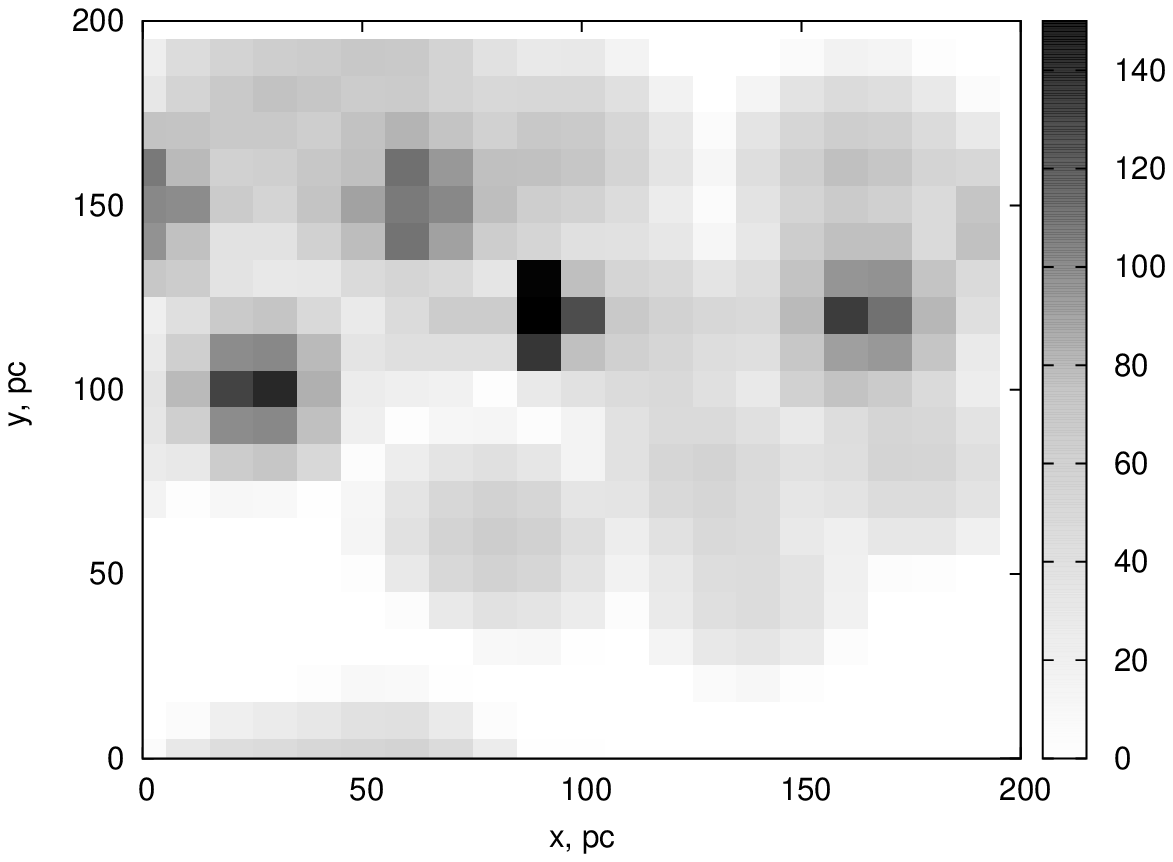,width=40mm,angle=0,clip=}
}
\break
\centerline{
\psfig{figure=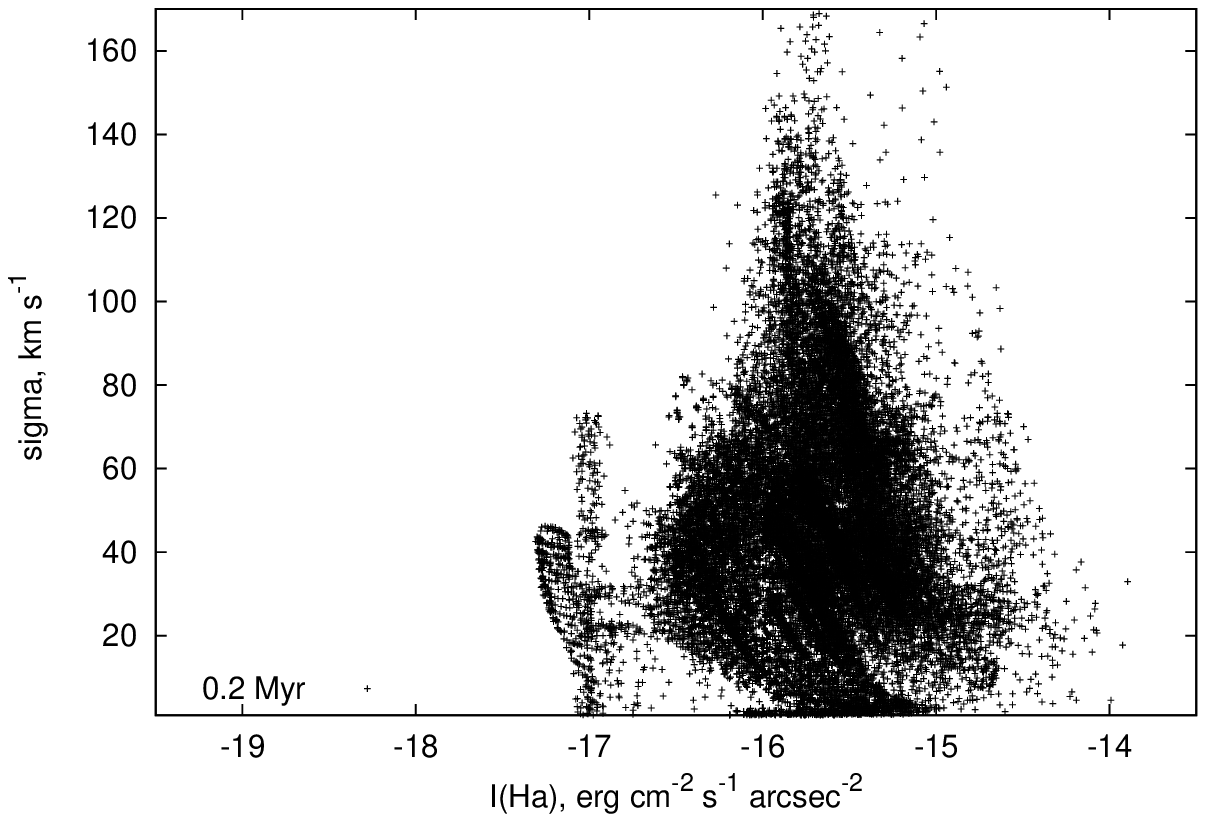,width=40mm,angle=0,clip=}
\psfig{figure=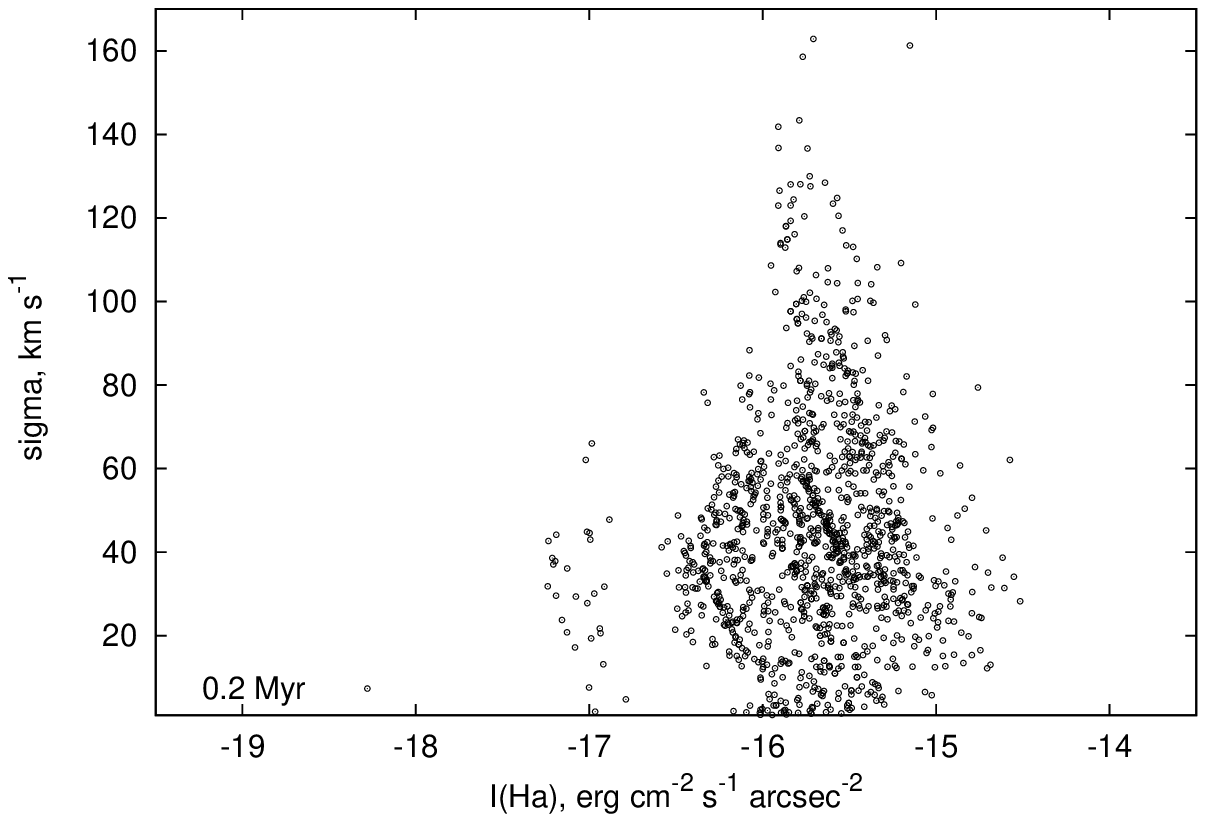,width=40mm,angle=0,clip=}
\psfig{figure=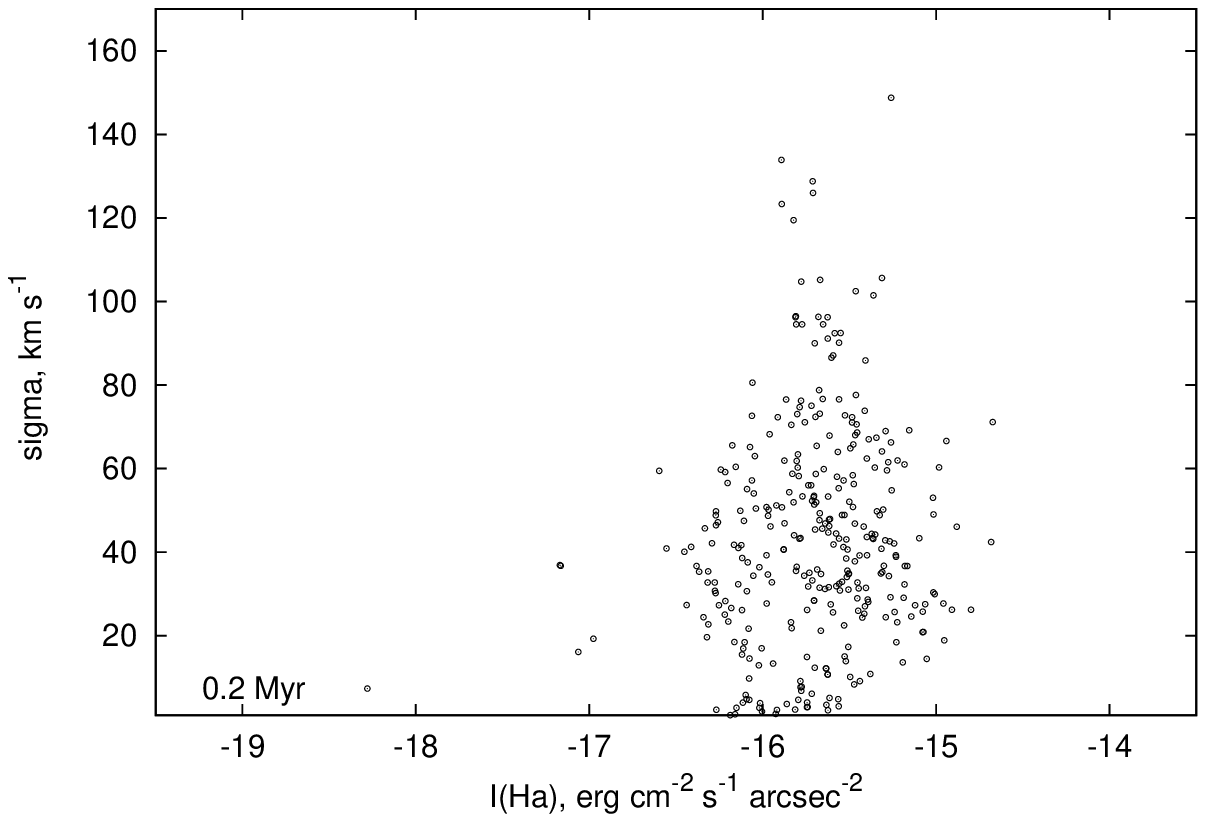,width=40mm,angle=0,clip=}
}
\vspace{1mm}
\captionb{4}
{
The H$\alpha$ intensity $I_{\rm H\alpha}$ (in erg~cm$^{-2}$~s$^{-1}$~arcsec$^{-2}$) -- top row, velocity dispersion 
(in km/s) maps -- middle row, and $I_{\rm H\alpha}-\sigma$ diagram -- bottom row, for SN rate $1.25\times
10^{-11}$~pc$^{-3}$~Myr$^{-1}$ at $t=0.2$~Myr. The numerical data without degrading resolution procedure is shown
in left column of panels, the smoothed and binned data are presented for $h=2.5$~pc and $N=5$~pc in middle column,
$h=5$~pc and $N=10$~pc in right column.
}
}
\label{ha-multisn}
\end{figure}
%%%%%%%%%%%%%%%%%%%%%%%%%%%%%%%%%%%%%%%%%%%%%%%%%%%%%%%%%%%%%%%%%%

Figure~4 (left column of panels) presents the H$\alpha$ intensity (upper), the velocity dispersion
(middle) maps and the $I_{\rm H\alpha}-\sigma$ diagram (bottom) for SN rate $1.25\times 10^{-11}$~pc$^{-3}$~Myr$^{-1}$ 
at $t=0.2$~Myr. Several SN bubbles are clearly identified (only two dozens SNe have exploded till this time), some of 
them interact between each other. It is clearly seen that high H$\alpha$ intensity and velocity dispersion are associated
with young SN remnants. But the highest velocity dispersion can be found around colliding shells of young and old SNe.
A breakthrough of high velocity gas of young SN shell into warm rarefied interiors of the old SN produces a strong shock
wave. So that we can conclude that collisions of SN shells with significant difference in age are resposible for velocity
dispersion reaches the value high as $\simgt 100$~km~s$^{-1}$. When two shells of SN with similar ages collide, a 
cold layer with divergent shocks is formed, in this case the velocity dispersion value does not increase significantly.
%%% Note that the velocity dispersion peak at high H$\alpha$ intensities corresponds to the most young SN remnant. 
Several peaks of $\sigma$ in the diagram conform to multiple stand-alone SN remnants with moderately different ages. 
Similar peaks can be found in dwarf galaxy DDO 53 (see Figure~1). For instance, similar picture 
takes place at $t\sim (3-5) \times 10^5$~yrs for SN explosions with frequency 1 per $10^5$~yr, whose shells do not (or
slightly) interact each other. 
%In some galaxies there are two strongly remarkable peaks with $\sigma \sim 100$~km~s$^{-1}$(like in IC 1613, see Figure~5 in Moiseev \& Lozinskaya, 2012). One can suppose that the peak with lower intensity and velocity dispersion is connected with older SN shells (some of them collide), whereas higher H$\alpha$ intensity peak corresponded to several compact regions is expected to be a signature of colliding shocks from recently merged SNe.Probably, a peak with similar $I_{\rm H\alpha} - \sigma$ relation appears due to the interaction of SN shock with stellar wind or with young HII shell around OB star.

%%%%%%%%%%%%%%%%%%%%%%%%%%%%%%%%%%%%%%%%%%%%%%%%%%%%%%%%%%%%%%%%%%
\begin{figure}[!tH]
\vbox{
\centerline{
\psfig{figure=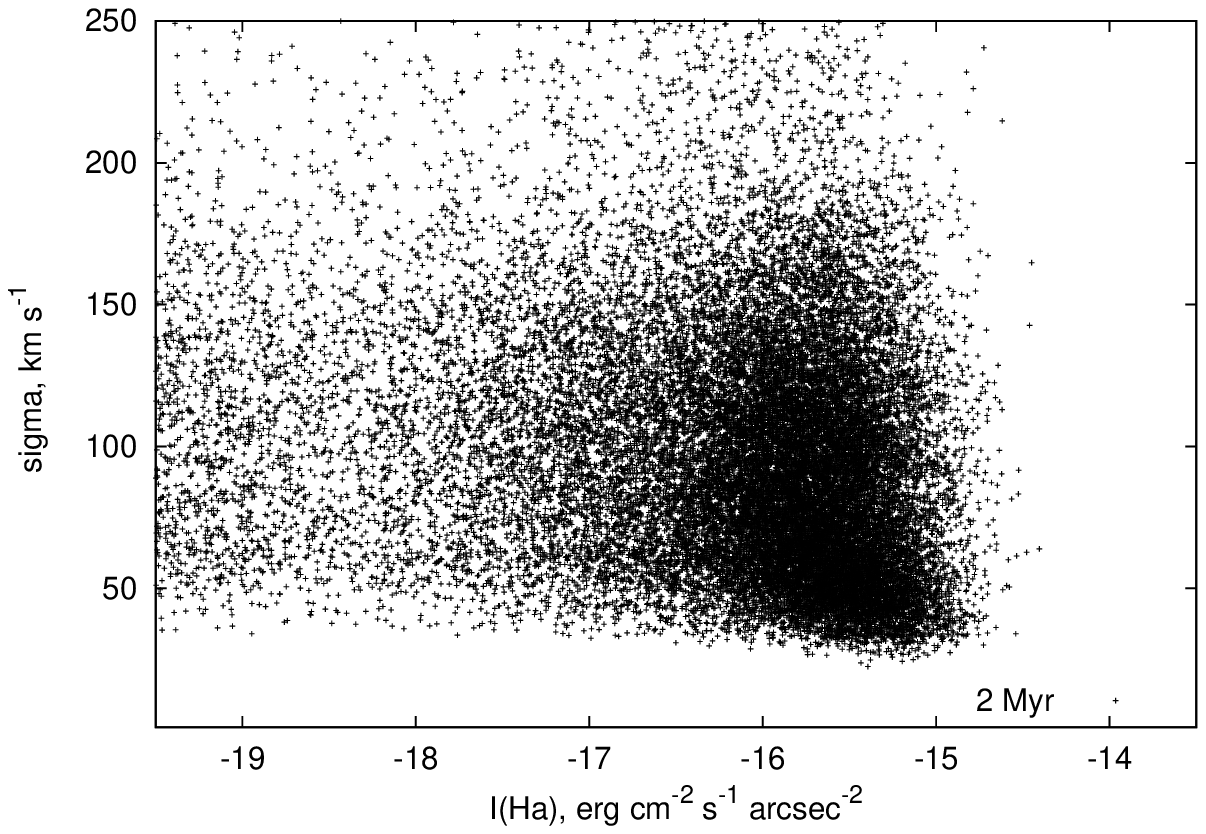,width=40mm,angle=0,clip=}
\psfig{figure=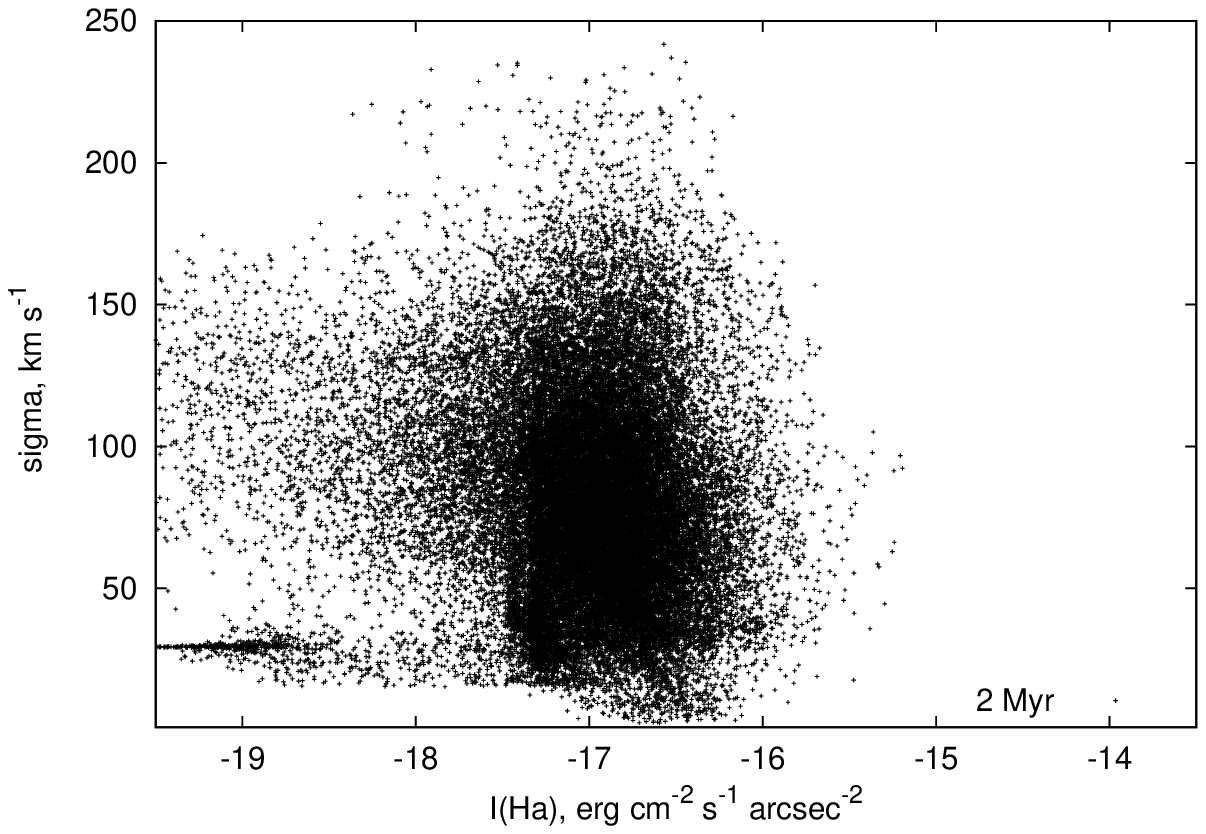,width=40mm,angle=0,clip=}
\psfig{figure=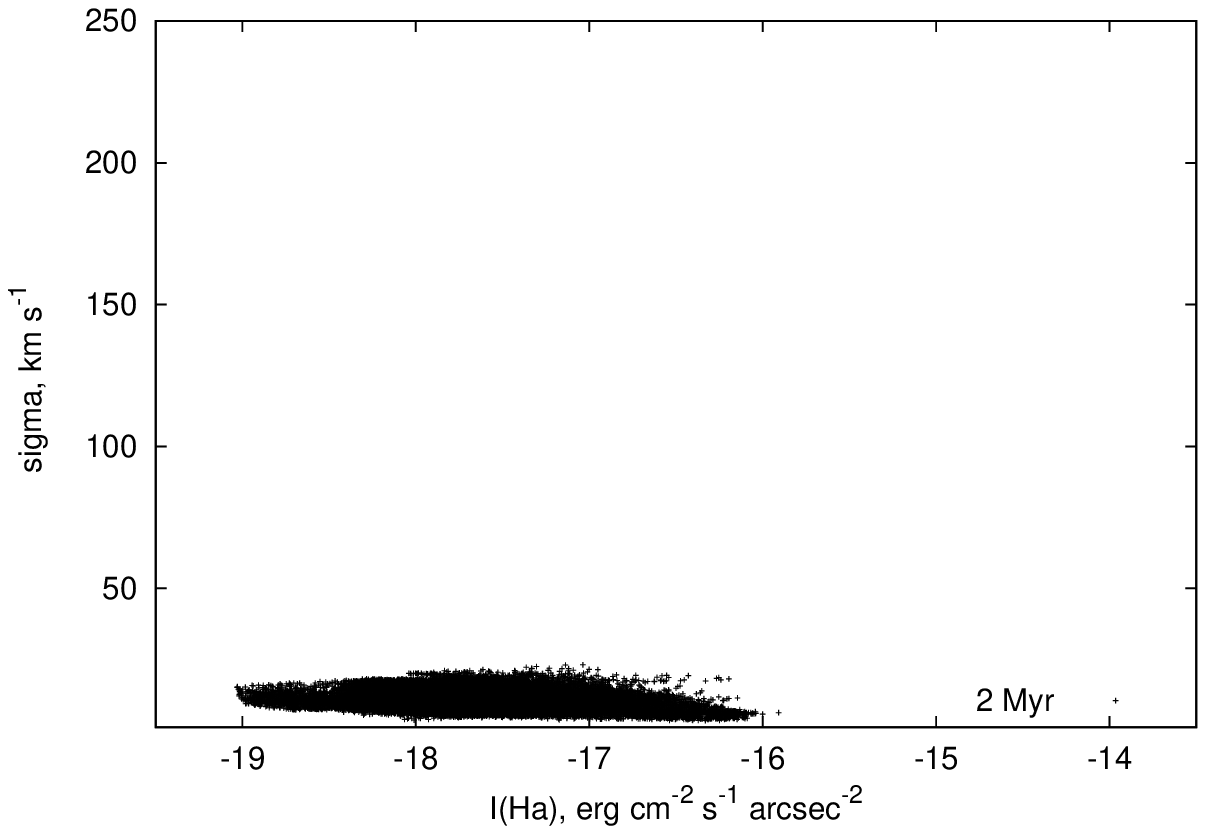,width=40mm,angle=0,clip=}
}
\vspace{1mm}
\captionb{5}
{
The $I_{\rm H\alpha}-\sigma$ diagrams for continous SN explosions with rate $1.25\times 10^{-11}$~pc$^{-3}$~Myr$^{-1}$ 
(1 SN per $10^4$~yrs in box 200~pc$^3$, left panel), $1.25\times 10^{-12}$~pc$^{-3}$~Myr$^{-1}$ (1 SN per $10^5$~yrs in 
box 200~pc$^3$, middle panel) and for limited (15) number SNe with delay $10^4$~yrs (right panel) at $t=2$~Myr.
}
}
\label{ha-last}
\end{figure}
%%%%%%%%%%%%%%%%%%%%%%%%%%%%%%%%%%%%%%%%%%%%%%%%%%%%%%%%%%%%%%%%%%

One can see that the $I_{\rm H\alpha}-\sigma$ diagram has a shape similar to observed one in dwarf galaxies after 10-20 SN
explosions (several of them should collide each other) with frequency 1 per $\sim 10^4$~yr. Decrease of frequency (or, in
general, SN rate) leads to smaller velocity dispersion. At further time the shape of $I_{\rm H\alpha}-\sigma$ diagram 
depends on whether SN exposions continue or stop. For the first case SN shell collisions lead to further increase of
velocity dispersion, which can reach $\sim 200-250$~km~s$^{-1}$ (Figure~5, left and middle panels). Both rarefied and hot 
gas has high velocity dispersion, but low H$\alpha$ intensity. Note that because the observed $\sigma$ maps were masked 
by the fixed S/N level (the left-hand side boundary of the point cloud in the observed diagrams is due to such masking), 
such rarefied and hot gas cannot be detected in observations. Decrease of SN frequency shifts the point cloud to lower
intensities (Figure~5, middle panel). For the case of limited SN number the SN shells lose their energy during $\sim 1$~Myr
(in gas with $n=0.1$~cm$^{-3}$) after the last SN has exploded, thus, both H$\alpha$ intensity and velocity dispersion 
drop to negligible values (Figure~5, right panel). 

Thus, the $I_{\rm H\alpha}-\sigma$ diagram obtained in our simulations is expected to be close to observed one, when
({\it i}) a starformation burst leads to 10-20 nearest SN explosions, ({\it ii}) several SN shells should collide, and 
({\it iii}) the next burst will be not earlier than previous SN remnants cool down and merge with ISM. Note that the 
diagram has a shape like in observation only in several hundred thousands years after the burst has finished, so
we can conclude that the last starformation episod in the observed galaxies, whose diagrams have similar shape, 
has taken place not earlier than several hundreds thousand years ago. 

%% commences, so we can conclude that the duration of the \textbf{last starformation episod(??????)}  in the observed galaxies, 
%% whose diagrams have similar shape, is not longer than several
%% hundreds thousand years. 

In observations the spatial resolution is much less than that in our simulations. To test our conclusions and compare 
our results with the observations we degrade the resolution of the H$\alpha$ and velocity dispersion maps. We smooth 
the maps with 2D Gaussian filter of $h$ size and after re-bin the smoothed maps with size of a bin equals to $N$. 
Figure~4 (middle and right columns of panels) presents the H$\alpha$ intensity and the velocity dispersion after the
degrading procedure. The maximum of velocity dispersion decreases from $\sim 160-170$~km~s$^{-1}$ in the original map 
to $\sim 120-140$~km~s$^{-1}$ for $h=5$~pc and $N=5$~pc (middle column) and to $\sim 100-110$~km~s$^{-1}$ for $h=10$~pc 
and $N=10$~pc (right column), this spatial resolution is close to that in the observations of IC~10. So the velocity 
dispersion peak in the degraded $I_{\rm H\alpha}-\sigma$  diagrams has a magnitude similar to that observed in dwarf 
galaxies (Mart\'inez-Delgado \etal 2007, Moiseev \etal 2010, Moiseev \& Lozinskaya 2012) and the model points occupy  
the similar ranges of the Balmer lines surface brightness and line-of-sight velocity dispersion.

\sectionb{4}{CONCLUSIONS}

We have studied the H$\alpha$ intensity -- velocity dispersion diagram for single and multiple SNe. We found that the
diagram obtained in our simulations are close in shape to those observed in the nearby dwarf galaxies. For single SN 
remnant the maximum velocity dispersion is about several dozens km~s$^{-1}$ and reaches it when the reverse shock is 
close to the blastwave. We concluded that collisions of SN shells with significant difference in age are resposible 
for high velocity dispersion that reaches values high as $\simgt 100$~km~s$^{-1}$. Peaks of velocity dispersion on the 
$I_{\rm H\alpha}-\sigma$ diagram may correspond to several stand-alone or merged SN remnants with moderately different 
ages. Similar structures can be found at the diagrams for dwarf galaxy DDO 53. We investigated the decrease 
of spatial resolution on shape of $I_{\rm H\alpha}-\sigma$ diagrams. We found that the $I_{\rm H\alpha}-\sigma$ 
diagrams obtained after the procedure of degrading resolution not only retain its similarity in shape, but also become
quantitatively close to those  observed dwarf galaxies.

\thanks{ 
EV and YuS thank to the RFBR for support (project codes 12-02-00365 and 12-02-00917).
EV and AM is also grateful for the financial support of the non-profit ``Dynasty'' Foundation. 
EV thanks to the Ministry of Education and Science of the Russian Federation (project 213.01-11/2014-5).
The work is partly supported by the Research Program OFN-17 of the Division of Physics, Russian Academy of Sciences.
}

\References

\refb Bordalo  V.,  Plana  H., Telles E., 2009, ApJ, 696, 1668

\refb Klingenberg Ch., Schmidt W., Waagan K., 2007, J. Comp. Phys., 227, 12

\refb Mart\'inez-Delgado I., Tenorio-Tagle G., Mu\~noz-Tu\~n\'on C., Moiseev A.V., Cair\'os L.M., 2007, AJ, 133, 2892

\refb Moiseev A.V., Lozinskaya T.A., 2012, MNRAS, 423, 1831

\refb Moiseev A.V., Pustilnik S.A., Kniazev A.Y., 2010, MNRAS, 405, 2453

\refb Mu\~noz-Tu\~n\'on C., Tenorio-Tagle G., Casta\~neda H.O., Terlevich R., 1996, AJ, 112, 1636

\refb Toro E., Riemann solvers and numerical methods for fluid dynamics. Springer-Verlag, Berlin, second edition, 1999. A practical introduction.

\refb Vasiliev E.O., 2011, MNRAS, 414, 3145

\refb Vasiliev E.O., 2012, MNRAS, 419, 3641

\refb Vasiliev E.O., 2013, MNRAS, 431, 638

\refb Vasiliev E.O., Nath B.B., Shchekinov Yu.A., 2014, MNRAS accepted, arXiv:1401.5070

\refb Yang H., Chu Y-H., Skillman E. D., Terlevich R., 1996, AJ, 112, 146

\end{document}